\documentclass[12pt]{article}

\usepackage{scicite}
\usepackage{xcolor}
\usepackage{times}
\usepackage{graphicx}

\newenvironment{sciabstract}{%
\begin{quote} \bf}
{\end{quote}}

\title{Direct determination of mode-projected electron-phonon coupling in the time-domain} 

\author{
M. X. Na$^{\ast,1,2}$, A. K. Mills$^{\ast,1,2}$, F. Boschini$^{1,2}$, M. Michiardi$^{1,2,3}$,\\ B. Nosarzewski$^{4}$, R. P. Day$^{1,2}$, E. Razzoli$^{1,2}$,\\ A. Sheyerman$^{1,2}$, M. Schneider$^{1,2}$, G. Levy$^{1,2}$, S. Zhdanovich$^{1,2}$,\\T. P. Devereaux$^{4}$, A. F. Kemper$^{5}$,  D. J. Jones$^{1,2\dagger}$, A. Damascelli$^{1,2\dagger}$\\
\\
\normalsize{$^{1}$Department of Physics and Astronomy, University of British columbia}\\
\normalsize{Vancouver, BC, V6T 1Z1, Canada}\\
\normalsize{$^{2}$Quantum Matter Institute, Vancouver, BC, V6T 1Z4, Canada}\\
\normalsize{$^{3}$Max Planck Institute for Chemical Physics of Solids, 01187 Dresden, Germany}\\
\normalsize{$^{4}$Department of Physics, Stanford University, Stanford, CA, 943305, USA}\\
\normalsize{$^{5}$Department of Physics, North Carolina State University, Raleigh, NC, 27695, USA}\\
\\
\normalsize{$^\ast$These authors equally contributed to the manuscript;}\\
\normalsize{$^\dagger$To whom correspondence should be addressed;}\\
\normalsize{E-mail:  djjones@physics.ubc.ca; damascelli@physics.ubc.ca}
}

\date{}

\begin{document} 


\baselineskip24pt

\maketitle 
\begin{sciabstract}
Ultrafast spectroscopies have become an important tool for elucidating the microscopic description and dynamical properties of quantum materials. In particular, by tracking the dynamics of non-thermal electrons, a material's dominant scattering processes \textemdash and thus the many-body interactions between electrons and collective excitations \textemdash can be revealed. Here we present a new method for extracting the electron-phonon coupling strength in the time domain, by means of time and angle-resolved photoemission spectroscopy (TR-ARPES). This method is demonstrated in graphite, where we investigate the dynamics of photo-injected electrons at the $\overline{\mathrm{K}}$ point, detecting quantized energy-loss processes that correspond to the emission of strongly-coupled optical phonons. We show that the observed characteristic timescale for spectral-weight-transfer mediated by phonon-scattering processes allows for the direct quantitative extraction of electron-phonon matrix elements, for specific modes, and with unprecedented sensitivity.
\end{sciabstract}

The concept of the electronic quasiparticle as proposed by Landau, \emph{i.e.} the dressing of an electron with many-body and collective excitations\cite{Landau}, is essential to the modern understanding of condensed matter physics. Among the plethora of interactions relevant to solid-state systems, electron-phonon coupling (EPC) has been a persistent subject of great interest, related as it is to the emergence of disparate physical phenomena, from resistivity in normal metals to conventional (BCS) superconductivity and charge-ordered phases \cite{LeTacon2013,Devereaux2016}. While strong EPC is desirable in systems like BCS superconductors\cite{Strongin1968, Allen}, it is deleterious for conductivity in normal metals, curtailing the application of several compounds as room-temperature electronic devices \cite{Scheuch2011b}.

Given the important role of the electron-phonon interaction in relation to both conventional and quantum materials, extensive theoretical and experimental efforts have been devoted towards determining the strength and anisotropy of EPC. While \emph{ab-initio} calculations are powerful, they rely on complex approximations which require precise experimental data to benchmark their validity\cite{Marsiglio2008}. Inelastic scattering experiments -- such as Raman spectroscopy \cite{Ferrari2007}, electron energy loss spectroscopy \cite{Tanaka2017}, inelastic x-ray \cite{Mohr2007a}, and neutron scattering \cite{Yildirim2001} -- are able to access EPC for specific phonon modes, yet integrated over all electronic states. Angle-resolved photoemission spectroscopy (ARPES), on the converse, can access the strength of EPC via phonon-mediated renormalization effects for specific momentum-resolved electronic states, as revealed by ``kinks" in the electronic band dispersion \cite{Valla1999,LaShell2000,Lanzara2001,Damascelli2003,Shi2004}. However, extraction of EPC strength from these kinks requires accurate modelling of the bare band dispersion and of the electronic self-energy, which can prove to be a formidable challenge either due to insufficient sensitivity and experimental resolutions\cite{Calandra}, or because of too-strong and/or compounded many-body interactions\cite{Ingle2005, Veenstra2011}. In addition, the interpretation of spectroscopic features is often elusive, as they may be attributed to several different many-body interactions\cite{Siegel 2012, Zhang2017, Li2018}.

Alternative and possibly more powerful approaches might come from the extension of ARPES into the time domain (TR-ARPES), which has already provided deep insights into the relaxation channels of hot electronic distributions, in which EPC plays a major role \cite{Perfetti2008, Ishida2011a,  Gierz2013c,  Sobota2014, Stange2015b, Yang2016}. On the one hand, TR-ARPES performed with 6~eV sources have enabled detailed study of low-energy many-body phenomena; while this has given a new perspective on superconducting gap dynamics of cuprate superconductors \cite{Perfetti2007, Smallwood2012a, Boschini2018}, electron-phonon interaction in bulk FeSe\cite{Gerber2017}, and surface-state dynamics in topological materials\cite{Sobota2012, Sobota2014}, low photon energies have limited these studies to a small region of the Brillouin Zone (BZ). On the other hand, ARPES systems using high-harmonic sources have extended the accessible momenta beyond the first BZ, but heretofore have focused on the high-energy scale electron dynamics on the order of 10 fs \cite{Rohwer2011, Gierz2013c, Stange2015b, Cilento}, as system energy resolutions have yet to reach the standards achieved by their 6~eV counterpart. 

In this work, we explore a new paradigm for the TR-ARPES study of transient spectral features at large momenta, made possible by a femtosecond high-harmonic source designed with specific emphasis on energy resolution\cite{Mills2018}. The experimental strategy is the following: we begin by injecting electrons into specific unoccupied states by optical excitation. As the hot electrons relax, we track specifically the transfer of spectral weight from these photo-excited states to lower-energy states via emission of a phonon with energy $\hbar\Omega_{\mathbf{q},\nu}$, where $\mathbf{q},\nu$ denote the phonon momentum and branch. The time constant extracted for this transfer of spectral weight ($\tau_{\mathbf{q},\nu}$) can then be directly related to the electron-phonon contribution to the self-energy for the phonon involved as\cite{Sentef2013}:
\begin{equation}
\frac{1}{\tau_{\mathbf{q},\nu}}=\frac{2\pi}{\hbar}\langle g_{\mathbf{q},\nu}^2\rangle D(E-\hbar\Omega_{\mathbf{q},\nu}),
\label{Eq. self-energy}
\end{equation}
where $E$ is the energy of the direct optical transition, $\langle g^2_{\mathbf{q},\nu}\rangle$ is the square of the mode-projected electron-phonon matrix element averaged over the states populated by optical excitation, and $D(E)$ is the electronic density of states (DOS)\cite{Sohier2016, Marsiglio2008} (derivation in Supplementary). We will show that this allows us to measure $\langle g^2_{\mathbf{q},\nu}\rangle$, gaining insight on the strength of the scattering process as well as the energy and momenta involved. 

\begin{figure}[p]
\centering
\includegraphics{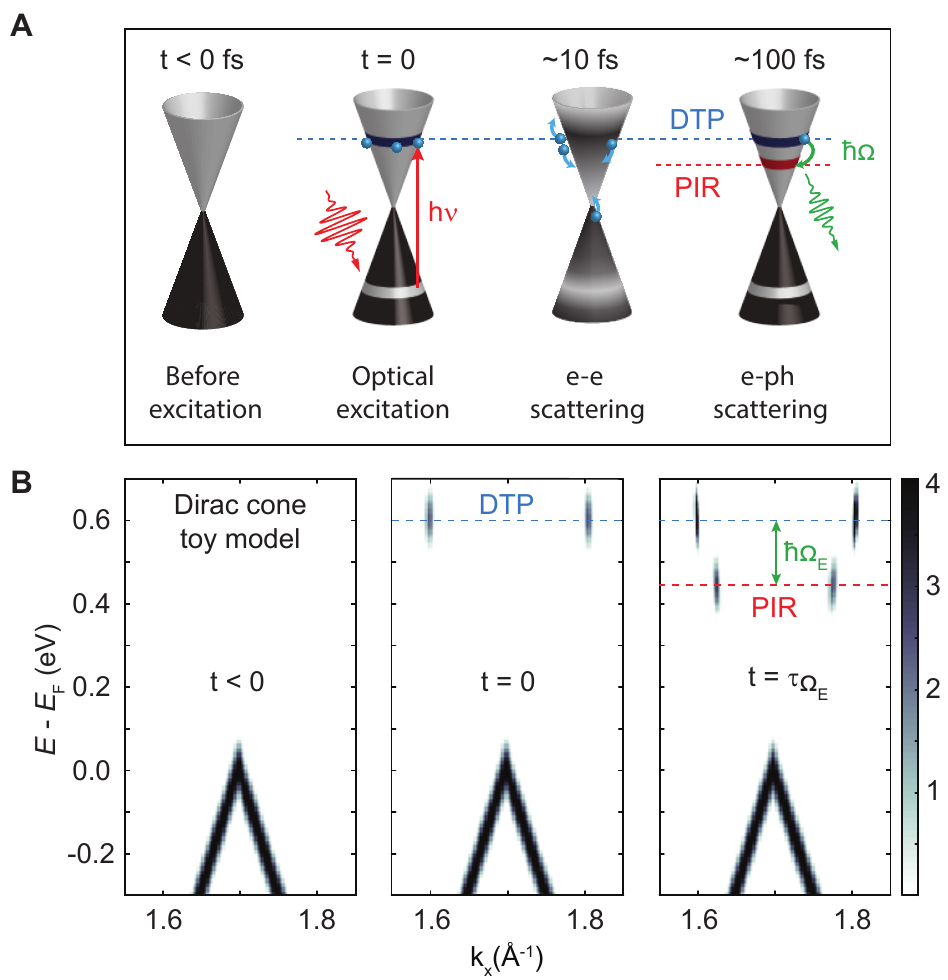}
\caption{\textbf{Toy model of optical injection and scattering processes on the Dirac cone.} (\textbf{A}) Sketch of the Dirac cone and electron dynamics. Black (gray) indicates occupied (unoccupied) states. During optical-excitation, electrons from the lower cone are promoted to the upper cone through a vertical transition (red arrow), creating a direct-transition peak (DTP). Electrons subsequently relax/scatter through electron-electron (e-e) and electron-phonon (e-ph) processes on timescales of 10~fs and 100~fs, respectively. The former (e-e) broadens the DTP, while the latter creates a phonon-induced replica (PIR) by the emission of a phonon. (\textbf{B}) Simulation of the transient TR-ARPES intensity for a Dirac cone pumped with 1.2~eV, including a retarded e-ph interaction with a phonon of energy $\hbar\Omega_{E}$. At time $t=0$, the DTP feature is observed at $E_{\Omega_E}=0.6$~eV; at $t=\tau_{\mathrm{ep}}$, the PIR is observed at $E_{\mathbf{DTP}}-\hbar\Omega_{E}$. The intensity of DTP and PIR features are enhanced ($\times 8$) for visualization purposes.}
\label{Fig1}
\end{figure}

In order to track the transfer of spectral weight, the initial (photo-injected) states and final states must be unambiguously defined and located. This is easiest on a small Fermi-surface, such as that of graphene, where phase-space scattering restrictions limit the number of initial and final states. To visualize the aforementioned electron-phonon scattering process, we simulate the pump-probe experiment using a Dirac dispersion as a toy model and calculate the transient ARPES spectra in response to optical excitation for the case of a single strongly-coupled Einstein phonon mode of energy $\hbar\Omega_{\mathbf{q},\nu}=\hbar\Omega_{\mathrm{E}}$. This model (Fig.\,\ref{Fig1}B) uses a two-time Green's function formalism on the Keldysh contour for a multi-orbital system. At $t=0$, the system is excited with a $1.2$~eV pulse, which promotes electrons via direct optical excitation to $0.6$~eV, observed experimentally as the direct-transition-peak (DTP). Then, on the characteristic timescale of $\tau_{\mathbf{q},\nu}=\tau_{\Omega_E}$,  the electron-phonon interaction leads to relaxation of the photo-injected electron population via the emission of phonons, resulting in the creation of a phonon-induced replica (PIR) at an energy $E_{\mathrm{PIR}}=0.6$~eV$-\hbar\Omega_E$. 

Here we perform the experiment on graphite, which has the same ideal phase-space restrictions as its monolayer counterpart, but without the need to consider coupling to the substrate, which in graphene is known to affect both electronic and phononic structure, as well as electron-phonon coupling\cite{Zhou2007,Wang2008, Allard2010}. The low-energy electronic structure of single-crystal graphite consists of two gapless nearly two-dimensional Dirac-like bands at the BZ corners (similar to graphene), as well as a second set of bands that disperses along the $c$-axis ($k_\perp$ in our experimental geometry, see Fig.\,\ref{Fig4}A)\cite{Slonczewski1958, Gruneis2008, Cheng2015}. In addition, graphite electrons are well-known to couple to optical phonons at $\overline{\Gamma}$ and $\overline{\mathrm{K}}$, \cite{Piscanec, Zhou2006, Mohr2007a, Lazzeri2008, Chatelain2014, Stern2018}, and have been extensively studied in both theory and experiment\cite{Ferrari2007, Ishida2011a, Yang2016}, making graphite an ideal benchmark system for the application of this new time-resolved technique.

\begin{figure}[p]
\centering
\includegraphics{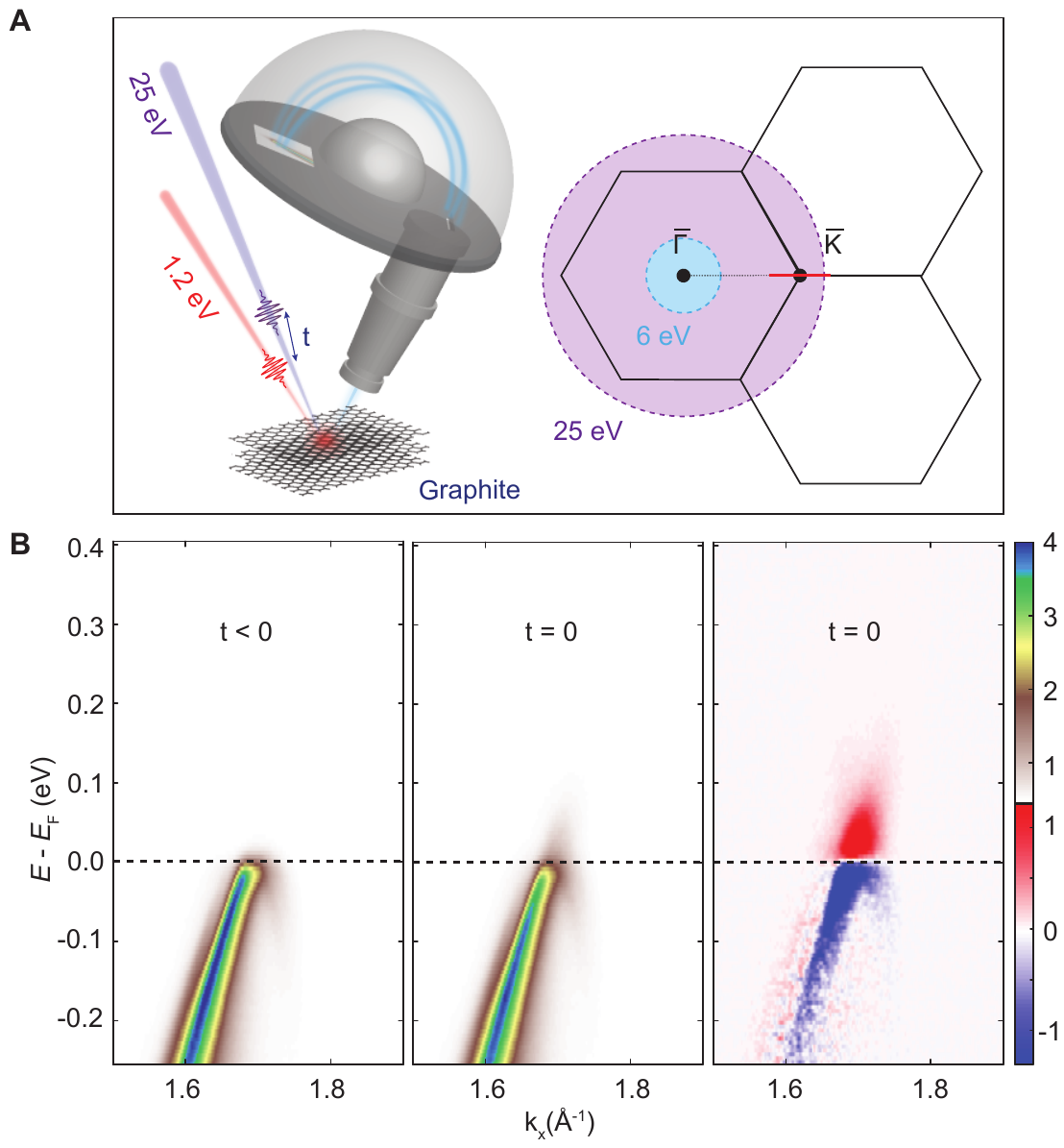}
\caption{\textbf{Electron dynamics measured by TR-ARPES in graphite.} (\textbf{A}) The experimental setup, along with the 2D-projected Brillouin Zone of graphite. Blue (purple) circles indicate the range of momenta accessible to 6 (25)~eV photons. We measure along the $\overline{\Gamma}-\overline{\mathrm{K}}$ direction, cut shown in red. (\textbf{B}) TR-ARPES measurements acquired with a 25 eV probe and 1.19 eV pump. The unpumped dispersion ($t<0$) is shown in the left panel. The pumped ARPES map at zero delay and its differential map are shown in the middle and right panels. Due to fast thermalization processes, the signal of DTP and PIR cannot be observed by simple visual inspection on a linear colormap.}  
\label{Fig2}
\end{figure}

Previous time-resolved experiments have shown that the time (energy) scale of the electron-phonon scattering process is on the order of 100~fs (100~meV). Therefore, observation of this transient spectral signature in TR-ARPES demands a balance of time and energy resolution. Achieving the system resolution requirements at photon energies needed to reach the $\overline{\mathrm{K}}$ point ($>$20~eV, assuming a maximum detection angle of 60 degrees; see Fig.\,\ref{Fig2}A for the BZ range covered by photons of different energy) was made viable by the development of a new cavity-based high-harmonic source\cite{Mills2018}. We select the 21$^{\mbox{st}}$ harmonic (25~eV) from the high-harmonic spectrum for photoemission, with an overall time (energy) resolution of 190~fs (22~meV), and a repetition rate of 60~MHz. The unpumped ARPES map of the Dirac-like dispersion along $\overline{\Gamma}$ - $\overline{\mathrm{K}}$ is shown in the left panel of Fig.\,\ref{Fig2}B, where only one branch of the cone is observed as a consequence of photoemission matrix elements\cite{Gierz2011,Liu2011}. In the middle and right panels, we show the pumped ARPES spectra at zero-delay, as well as its differential map (obtained by subtracting the equilibrium map from its counterpart at zero pump-probe delay). The data were measured under perturbative excitation by a 1.19~eV pump pulse with an incident fluence of 18~$\mu$J/cm$^2$, where red (blue) color indicate a transient increase (decrease) of photoemission intensity. The timescales of the anticipated primary scattering processes following optical excitation are sketched in  Fig.\,\ref{Fig1}A. Following creation of the DTP above $E_F$, electrons decay into a thermal distribution, mainly via electron-electron (e-e) and electron-phonon (e-ph) scattering events. Since the e-e scattering processes are about an order of magnitude faster than e-ph scattering\cite{Ulstrup}, we should observe a rapid build up of photoemission intensity at the Fermi energy. This can render the observation of the DTP and PIR non-trivial, requiring a careful analysis of features above the hot electron background.

\begin{figure}[p]
\centering
\includegraphics{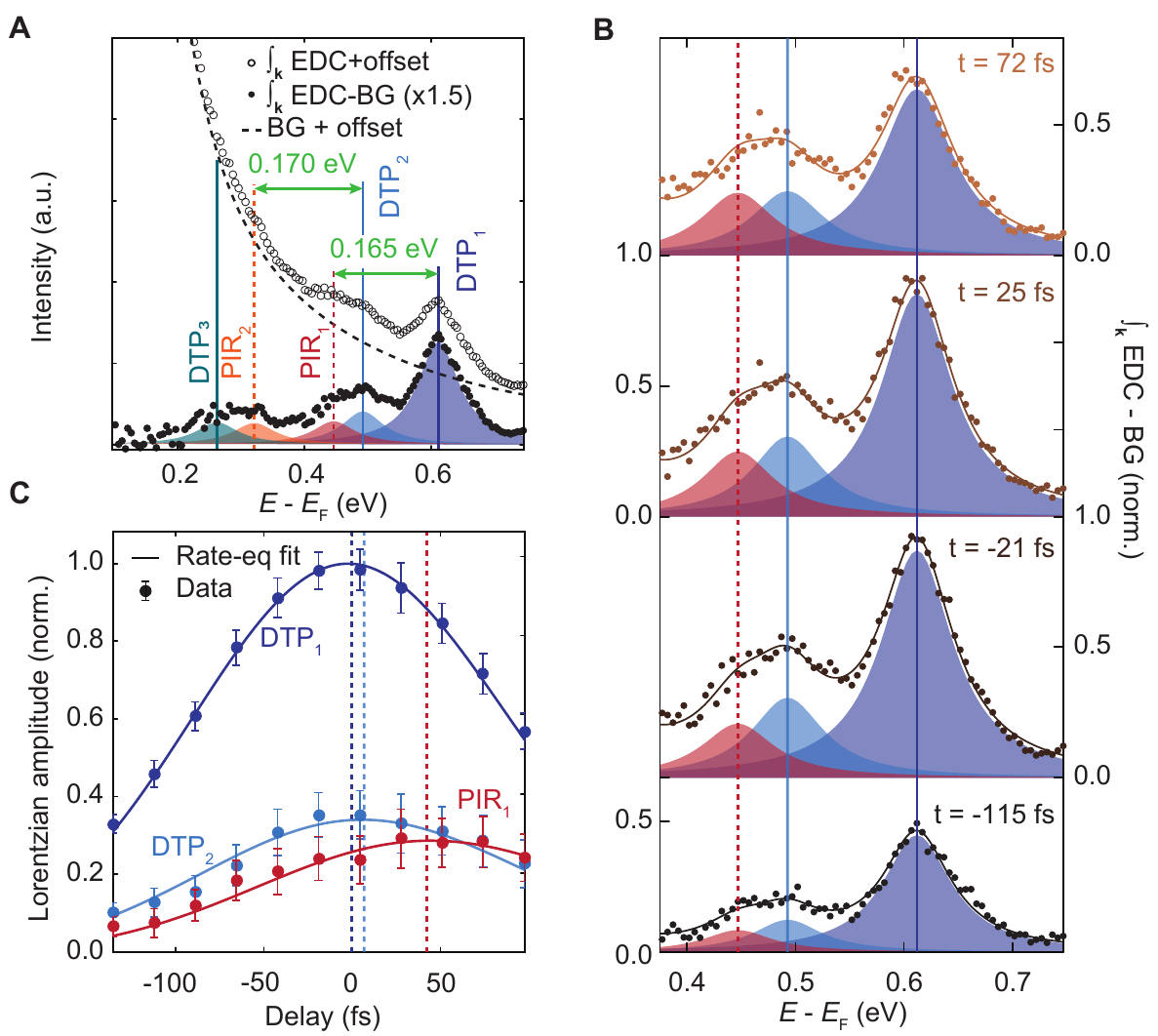}
\caption{\textbf{Time dependence of photo-induced excitations in graphite.} (\textbf{A}) Open circles display the momentum-integrated energy distribution curve ($\int_{\mathbf{k}}$ EDC), where signal is integrated in momentum along the  $\overline{\Gamma}$-$\overline{\mathrm{K}}$ direction. The subtraction of the bi-exponential hot-electron background (BG) highlights a series of peaks (filled circles), which are a combination of DTPs and PIRs. Solid (dashed) lines indicate the fitted position of DTP (PIR) peaks. The phonon energies extracted between the DTP$_1$/PIR$_1$ and DTP$_2$/PIR$_2$ pairs are 0.165~eV and 0.170~eV respectively, as indicated by the green arrow. (\textbf{B}) Evolution of the most prominent peaks. Dark (light) blue correspond to DTP$_1$ (DTP$_2$), red corresponds to PIR$_1$. The amplitudes are indicative of the population of electrons in each state. (\textbf{C}) The amplitude of the Lorentzians for each peak shown in B are plotted as a function of time. Dashed lines indicate the peak delay: DTP$_2$ (PIR$_1$) is delayed 13 (44)~fs with respect to DTP$_1$. Solid lines are the electronic occupation for the specified states derived from the rate-equation model fit. The transfer of spectral weight from DTP$_1$ to PIR$_1$ is associated to an e-ph scattering time constant $\tau_{\mathbf{q},\nu}=174\pm 35$~fs.}  
\label{Fig3}
\end{figure}

\begin{figure}[p]
\includegraphics[width=1\columnwidth]{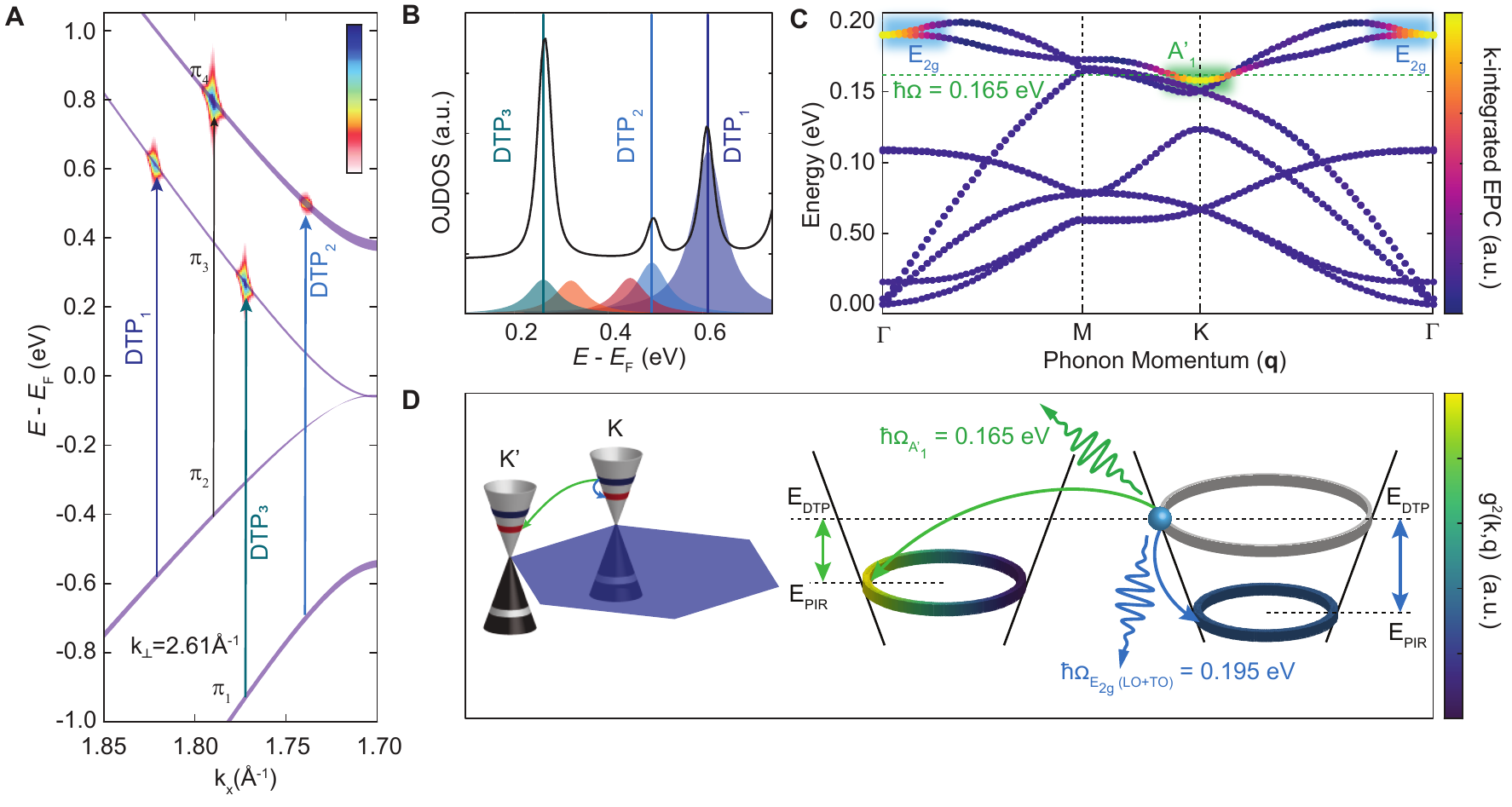}
\caption{\textbf{Calculation of the optical-joint-DOS and e-ph coupling.} (\textbf{A}) Momentum-resolved optical-joint-DOS, extracted using a modified tight-binding model from Ref\cite{Cheng2015}, showing the available optical transitions between the bands of graphite for a pump photon energy of 1.19~eV, integrated around $k_\perp=2.61\mathrm{\AA}^{-1}$. This value is fitted using an inner potential of $V_0=16.4\pm 0.1$~eV\cite{Gruneis2008}. (\textbf{B}) Integrated optical-joint-DOS along the $\overline{\Gamma}-\overline{\mathrm{K}}$ direction. The peaks extracted from panel A in Fig.\,\ref{Fig3} are also shown, with the location of the fitted DTP overlaid in solid lines, displaying good agreement with the optical-joint-DOS. (\textbf{C}) The phonon-dispersion of graphite as calculated from DFT\cite{Sohier2016}. Colors represent the strength of the total EPC (i.e. integrated over all electronic momenta). The dominant modes $A_1'$ ($E_{2g}$) are highlighted in green (blue)\cite{Piscanec}. (\textbf{D}), Calculation of the $\langle g^2_{\mathbf{k},\mathbf{q}}\rangle$. Case one: electron scattering with an $A_1'$ mode (green) with momentum $\sim\mathbf{K}$ from a specific state (indicated by blue sphere) in the 0.61 eV energy contour at K to states in the 0.45 eV contour at K'. Case two: electron scattering with an $E_{2g}$ (blue) mode with momentum $\sim 0$ from a specific state in the 0.61 eV energy contour at K to states in the 0.41 eV contour at K. $\langle g^2_{A_1'}\rangle$  and $\langle g^2_{E_{2g}}\rangle$ are obtained by integrating over both the constant energy contour at $E_{\mathrm{PIR}}$ and the electron position along the contour at $E_{\mathrm{DTP}}$. Colors indicate that the value of $\langle g^2_{A_1'}\rangle$ is twice that of the LO+TO combined $\langle g^2_{E_{2g}}\rangle$(see text for precise values). The scattering process from K' to K (not shown) is identical.}
\label{Fig4}
\end{figure}

In Fig.\,\ref{Fig3}A (black open circles) we display the momentum-integrated energy distribution curve ($\int_\mathbf{k}$~EDC) along the $\overline{\Gamma}$-$\overline{\mathrm{K}}$  direction.  We stress that the $\int_\mathbf{k}$~EDC is proportional to the occupied DOS along the selected momentum cut shown in Fig.\,\ref{Fig2}A (see Supplementary Information). Black filled circles in Fig.\,\ref{Fig3}A represent the $\int_\mathbf{k}$~EDC after removal of a bi-exponential background given by the thermal electronic distribution (near $E_F$) and non-thermal e-e scattering processes (near 0.6~eV). Once this background is removed, the $\int_\mathbf{k}$~EDC directly exposes the transient peaks, which can be fitted with five Lorentzians of the same width (Fig.\,\ref{Fig3}A). We can immediately identify the prominent peak at 0.6~eV as DTP$_1$, which was anticipated in the toy-model (Fig.\,\ref{Fig1}B), and associated with the optical transition from the $\pi_2$-to-$\pi_3$ band in Fig.\,\ref{Fig4}A. The other peaks -- as we will illustrate in more detail below -- are a combination PIRs and other DTPs, which arise from the second set of electronic bands ($\pi_1, \pi_4$) that disperse in $k_\perp$. We confirm these transitions in Fig.\,\ref{Fig4}A with a calculation of the optical-joint-DOS for graphite, adapted from a tight-binding model in Ref\cite{Cheng2015}, for a pump photon energy of 1.19~eV. The possible transitions along the $\overline{\Gamma}-\overline{\mathrm{K}}$ cut are shown in Fig.\,\ref{Fig4}A. While the $\pi_2$-to-$\pi_4$ transition is outside the range of our data, the three lower DTPs fall exactly in the energy range we expect. The resulting momentum-integrated optical-joint-DOS is shown in Fig.\,\ref{Fig4}B, along with the energy position of the five fitted peaks from Fig.\,\ref{Fig3}A.

To illustrate the DTP-to-PIR scattering process, we focus on the time-evolution of the three most prominent peaks, shown in Fig.\,\ref{Fig3}B, and relegate discussion of the DTP$_2$/PIR$_2$ pair to the Supplementary Information. The combined time and energy resolution of our source allows for a detailed study of the transient evolution of the DTP and the PIR, given by the amplitude of the Lorentzians in Fig.\,\ref{Fig3}C. Despite being only 50~meV apart, the dynamics of the light-blue and red peaks are markedly different. The population of the light-blue peak is only slightly delayed with respect to the dark-blue DTP$_1$, and is identified with a direct optical excitation ($\pi_1$-to-$\pi_4$ band in Fig.\,\ref{Fig4}A, labelled DTP$_2$), with the temporal delay being a consequence of energy-dependent electron lifetime \cite{Xu1996, Narang2017}. In contrast, the population of the red peak is delayed by $\Delta t=44\pm 15$~fs, too large to be compatible with optical excitation. It instead corresponds to the simulated PIR in Fig.\,\ref{Fig1}B, where the energy of the phonon involved is $\hbar\Omega_{\mathbf{q},\nu} = E_{\mathrm{DTP}}-E_{\mathrm{PIR}}=0.165\pm 0.011$~eV. 

The solid lines in Fig.\,\ref{Fig3}C is the result of a phenomenological rate-equation model describing the transfer of spectral weight between the DTP and the PIR pairs (see Supplementary for details). The population of electrons in the (dark/light) blue DTP are governed by rate-equations involving three terms: population by a 120~fs pump pulse, energy-dependent thermalization of the excited state population to the hot electron bath ($\tau_{\mathrm{th}}$), and energy-dependent phonon-mediated decay of the excited state population ($\tau_{\mathbf{q},\nu}$). This latter term transfers spectral weight from the DTP$_i$ to the PIR$_i$, which lose electrons to the same thermalization and phonon-mediated decay terms. The resultant temporal evolution is then convolved with a Gaussian with a full-width-at-half-maximum of 150~fs to account for the pulse duration of the photoemission probe. With this simple model, we find that a thermalization constant of $\tau_{\mathrm{th}}=56\pm 16$~fs and an e-ph decay constant of $\tau_{\mathbf{q},\nu}=174\pm 35$~fs well reproduce the delay and relative population of the non-thermal signatures. In addition, the extracted $\tau_{\mathbf{q},\nu}$ is consistent with \textit{ab-initio} calculations and estimates for e-ph scattering time in previous time-resolved studies\cite{Ulstrup, Stange2015b,Johannsen2013, Gierz2013c, Yang2016}. Since we are coupling to a single phonon mode of energy $\hbar\Omega_{\mathbf{q},\nu}=0.165\pm 0.011$~eV, we can directly relate this time constant to the mode-projected e-ph matrix element via Eq.\,\ref{Eq. self-energy}, using an electronic DOS [$D(E-\hbar\Omega_{\mathbf{q},\nu})=0.0241$ (1/eV)] calculated from the tight-binding model in Fig.\,\ref{Fig4}A. From this, we obtain a value of $\langle g_{\mathbf{q},\nu}^2\rangle=0.050\pm 0.011~\mathrm{eV}^2$.

We now assign the observed PIR to scattering by a specific phonon mode by comparing the extracted $\hbar\Omega_{\mathbf{q},\nu}$ against the phonon-dispersion of graphite calculated by density functional theory (DFT) in Fig.\,\ref{Fig4}C. The colors represent the EPC integrated over all electronic momenta, and indicates strong coupling for the $E_{2g}$ mode at $\mathrm{\Gamma}$ and the $A_1'$ mode at $\mathrm{K}$. The latter is the phonon mode associated with DTP/PIR pair we observe, as its energy matches the 0.165~eV we extract (green dashed line). Given that it has momentum $\mathrm{K}$, this mode is associated with the intervalley scattering of electrons between states at $\overline{\mathrm{K}}$ and $\overline{\mathrm{K'}}$. We consider this scattering process explicitly for a single electron in Fig.\,\ref{Fig4}D (green arrow). Starting from an initial state $i$ on the constant energy contour $E_{\mathrm{DTP}_{1}}$, we calculate the matrix element $g^2_{\mathbf{k},\mathbf{q}}$ for scattering events leading to the final state $f$ on the constant energy contour $E_{\mathrm{DTP}_{1}}-\hbar\Omega_{A_1'}$, such that $\mathbf{k}_f-\mathbf{k}_i=\mathbf{q}$ is fulfilled. This mode-projected calculation gives a value of $\langle g_{A_1'}^2\rangle=0.040$~eV$^2$, in remarkable agreement with the experimental value of $\langle g^2_{\mathbf{q},\nu}\rangle=\langle g_{A_1'}^2\rangle=0.050\pm 0.011$~eV$^2$ we previously extracted from the rate-equation fits to the experimental data. In  addition to the $A_1'$ mode, Fig.\,\ref{Fig4}C suggests that the $\Gamma-E_{2g}$ modes (LO and TO) are also expected to be strongly-coupled; however, considering the scattering process as before, we extract for the degenerate LO and TO modes a total coupling $\langle g_{E_{2g}}^2\rangle=0.023$~eV$^2$, which corresponds to a time constant of $ > 300$~fs. This coupling is approximately $50\%$ that of the $A_1'$ mode, consistent with previous theoretical considerations \cite{Butscher2007}. Thus, the PIR associated with emission of the $\Gamma-E_{2\mathrm{g}}$ phonons would not be visible above the hot electron background in our experiment.

In this work, we have demonstrated that TR-ARPES is capable of identifying phonon modes involved in electron relaxation processes, and can access the mode-projected e-ph matrix element through the dynamics of non-thermal spectral features. In particular, using this method, we have shown a proof-of-principle extraction of the mode-projected e-ph matrix element $\langle g^2_{A_1'}\rangle$ in graphite. We remark that $\langle g^2\rangle$ is a fundamental quantity, defined as the momentum average of the change in the electronic Hamiltonian in response to the ionic displacements of a phonon (see Supplementary, and Eq.\,S.\,13). In particular, $\langle g^2\rangle$ is independent of  doping and specified for a well-defined set of initial and final electronic states, and a well-defined bosonic mode (in this case, the $A_1'$ optical phonon with momentum $\mathbf{K}$ in graphite). Nonetheless, it is instructive to estimate the EPC constant $\lambda=2\langle g^2\rangle D(E_F)/(\hbar\Omega)$ for comparison with other approaches\cite{Marsiglio2008}. Commonly seen in relation to the critical temperature in superconductivity, the quantity $\lambda$ is doping dependent, and integrated over all bosonic and electronic degrees of freedom. Therefore, caution must be applied in comparing the two quantities. In pristine and low doping graphene/graphite systems, the vanishing DOS at the Dirac-point of graphene (crossing point of graphite) makes extraction of $\lambda$ very difficult, with reported values ranging from $4\times 10^{-4}$ to $1.1$\cite{Sugawara, Zhou2006, Bostwick2007, Leem2008, Siegel2012, Ulstrup2012, Joucken2016}, while DFT predicts $\lambda<0.05$\cite{Calandra}. Even forgoing doping dependence, this clearly illustrates the difficulty ARPES has in extracting the EPC of this particular system. In this work, we extract $\langle g^2_{A_1'}\rangle$ for the DTP state at $0.6$~eV, corresponding to a value of the mode-projected EPC $\lambda_{A_1'}=0.0182\pm 0.004$. This value characterizes a system where the Fermi-level is doped up to $0.6$~eV above the crossing point (the value at zero-doping would be $\lambda_{A_1'}=0.006\pm 0.001$). While we would ideally compare this value to that extracted by kink-analysis in graphite, the strong curvature of the bare-band dispersion\cite{Leem2008} and the lack of studies at comparable doping make this particularly challenging. Thus, we compare this value to what is reported in Ref.\,\cite{Siegel2012}, a doping-dependent study of $\lambda$ in graphene which is additionally supported by DFT calculations\cite{Calandra}. When the system is doped such that the Dirac point is 0.6~eV below $E_F$ [corresponding to a carrier density of $n\approx 4\times 10^{13}~(1/\mathrm{cm}^2)$], then $\lambda\approx 0.035$ is extracted. From this, we see that $\lambda_{A_1'}\approx \lambda/2$, which is consistent with the fact that $\lambda_{A_1'}$ captures only one of two strongly coupled modes in the system (the other being $E_{2g}$), while $\lambda$ extracted from kink-analysis is integrated over all modes. Altogether, these results suggest that time-domain measurements have the ability to access the EPC in a precise, sensitive, and mode-projected way. 

In principle, this technique is applicable to materials in which electrons are sufficiently strongly-coupled to one or few bosonic modes, such that distinct boson-induced replicas can be observed. Beyond graphite and graphene, quasi-two-dimensional materials such as transition-metal dichalcogenides feature gapped bands at the K and K' points, with further restrictions of the phase space for scattering stemming from valley degrees of freedom, and would be ideal candidates for similar TR-ARPES studies.  In addition, also conventional and unconventional superconductors, such as MgB$_2$ or cuprate/Fe-based superconductors, respectively, famously feature strong coupling to bosonic modes, which may drive electronic renormalizations (kinks). By monitoring quantized decay processes across the full BZ, this non-equilibrium technique will offer a new approach for studying the microscopic origin and momentum dependence of electron-boson coupling and its role in the emergence of superconductivity. 

These results also prove that a new analytical perspective can be achieved in TR-ARPES by taking advantage of novel femtosecond sources that combine high-photon energy (to access large electronic momenta) with high energy resolution (to resolve the low-energy quasiparticle dynamics). With the development of tunable pumps, polarization control for pump and probe, and bandwidth control to balance the tradeoff between energy and time resolution, an unprecedented versatility will be available for TR-ARPES experiments. By monitoring the population of electrons injected into momentum- and energy- selected states by direct optical excitation, one could formulate a series of new studies on empty state dispersion, lifetime, decoherence, and electron-boson interactions in a wide range of quantum materials.

We gratefully acknowledge M. Berciu, G. A. Sawatzky, A. Nocera, Z. Ye and D. Manske for critical reading of the manuscript and useful discussions, as well as C. Gutierr\' ez for figure design. This research was undertaken thanks in part to funding from the Max Planck-UBC-UTokyo Centre for Quantum Materials and the Canada First Research Excellence Fund, Quantum Materials and Future Technologies Program. The work at UBC was supported by the Gordon and Betty Moore Foundation's EPiQS Initiative, Grant GBMF4779, the Killam, Alfred P. Sloan, and Natural Sciences and Engineering Research Council of Canada's (NSERC's) Steacie Memorial Fellowships (A.D.), the Alexander von Humboldt Fellowship (A.D.), the Canada Research Chairs Program (A.D.), NSERC, Canada Foundation for Innovation (CFI), British Columbia Knowledge Development Fund (BCKDF), and the CIFAR Quantum Materials Program. E.R. acknowledges support from the Swiss National Science Foundation (SNSF) grant no. P300P2\_ 164649. A.F.K. acknowledges support by the National Science Foundation under Grant DMR-1752713.

\pagebreak


\begin{thebibliography}{10}

\bibitem{Landau}
L.~D. Landau, {\it J. Exp. Theor. Phys.\/} {\bf 3}, 920 (1957).

\bibitem{LeTacon2013}
M.~{Le Tacon}, {\it et~al.\/}, {\it Nat. Phys.\/} {\bf 10}, 52 (2014).

\bibitem{Devereaux2016}
T.~P. Devereaux, {\it et~al.\/}, {\it Phys. Rev. X\/} {\bf 6}, 041019 (2016).

\bibitem{Strongin1968}
M.~Strongin, {\it et~al.\/}, {\it Phys. Rev. Lett.\/} {\bf 21}, 1320 (1968).

\bibitem{Allen}
W.~L. McMillan, {\it Phys. Rev.\/} {\bf 167}, 331 (1968).

\bibitem{Scheuch2011b}
M.~Scheuch, {\it et~al.\/}, {\it Appl. Phys. Lett.\/} {\bf 99}, 211908 (2011).

\bibitem{Marsiglio2008}
F.~Marsiglio, J.~P. Carbotte, {\it Electron-Phonon Superconductivity\/}
  (Springer Berlin Heidelberg, Berlin, Heidelberg, 2008), pp. 73--162.

\bibitem{Ferrari2007}
A.~C. Ferrari, {\it Solid State Commun.\/} {\bf 143}, 47 (2007).

\bibitem{Tanaka2017}
S.-i. Tanaka, K.~Mukai, J.~Yoshinobu, {\it Phys. Rev. B\/} {\bf 95}, 165408
  (2017).

\bibitem{Mohr2007a}
M.~Mohr, {\it et~al.\/}, {\it Phys. Rev. B\/} {\bf 76}, 035439 (2007).

\bibitem{Yildirim2001}
T.~Yildirim, {\it et~al.\/}, {\it Phys. Rev. Lett.\/} {\bf 87}, 37001 (2001).

\bibitem{Valla1999}
T.~Valla, A.~V. Fedorov, P.~D. Johnson, S.~L. Hulbert, {\it Phys. Rev. Lett.\/}
  {\bf 83}, 2085 (1999).

\bibitem{LaShell2000}
S.~LaShell, E.~Jensen, T.~Balasubramanian, {\it Phys. Rev. B\/} {\bf 61}, 2371
  (2000).

\bibitem{Lanzara2001}
A.~Lanzara, {\it et~al.\/}, {\it Nature\/} {\bf 412}, 510 (2001).

\bibitem{Damascelli2003}
A.~Damascelli, {\it Phys. Scr.\/} {\bf T109}, 61 (2004).

\bibitem{Shi2004}
J.~Shi, {\it et~al.\/}, {\it Phys. Rev. Lett.\/} {\bf 92}, 186401 (2004).

\bibitem{Calandra}
M.~Calandra, F.~Mauri, {\it Phys. Rev. B\/} {\bf 76}, 205411 (2007).

\bibitem{Ingle2005}
N.~J. Ingle, {\it et~al.\/}, {\it Physical Review B\/} {\bf 72}, 205114 (2005).

\bibitem{Veenstra2011}
C.~N. Veenstra, G.~L. Goodvin, M.~Berciu, A.~Damascelli, {\it Physical Review
  B\/} {\bf 84}, 085126 (2011).

\bibitem{Siegel2012}
D.~A. Siegel, C.~Hwang, A.~V. Fedorov, A.~Lanzara, {\it New J. Phys.\/} {\bf
  14} (2012).

\bibitem{Zhang2017}
C.~Zhang, {\it et~al.\/}, {\it Nat. Commun.\/} {\bf 8}, 14468 (2017).

\bibitem{Li2018}
F.~Li, G.~A. Sawatzky, {\it Physical Review Letters\/} {\bf 120}, 237001
  (2018).

\bibitem{Perfetti2008}
L.~Perfetti, {\it et~al.\/}, {\it New J. Phys.\/} {\bf 10}, 053019 (2008).

\bibitem{Ishida2011a}
Y.~Ishida, {\it et~al.\/}, {\it Sci. Rep.\/} {\bf 1}, 64 (2011).

\bibitem{Gierz2013c}
I.~Gierz, {\it et~al.\/}, {\it Nat. Mater.\/} {\bf 12}, 1119 (2013).

\bibitem{Sobota2014}
J.~A. Sobota, {\it et~al.\/}, {\it Journal of Electron Spectroscopy and Related
  Phenomena\/} {\bf 195}, 249 (2014).

\bibitem{Stange2015b}
A.~Stange, {\it et~al.\/}, {\it Phys. Rev. B\/} {\bf 92}, 184303 (2015).

\bibitem{Yang2016}
J.~A. Yang, S.~Parham, D.~Dessau, D.~Reznik, {\it Sci. Rep.\/} {\bf 7}, 40876
  (2017).

\bibitem{Perfetti2007}
L.~Perfetti, {\it et~al.\/}, {\it Phys. Rev. Lett.\/} {\bf 99}, 197001 (2007).

\bibitem{Smallwood2012a}
C.~L. Smallwood, {\it et~al.\/}, {\it Science\/} {\bf 336}, 1137 (2012).

\bibitem{Boschini2018}
F.~Boschini, {\it et~al.\/}, {\it Nat. Mater.\/} {\bf 17}, 416 (2018).

\bibitem{Gerber2017}
S.~Gerber, {\it et~al.\/}, {\it Science\/} {\bf 357}, 71 (2017).

\bibitem{Sobota2012}
J.~A. Sobota, {\it et~al.\/}, {\it Phys. Rev. Lett.\/} {\bf 108}, 117403
  (2012).

\bibitem{Rohwer2011}
T.~Rohwer, {\it et~al.\/}, {\it Nature\/} {\bf 471}, 490 (2011).

\bibitem{Cilento}
F.~Cilento, {\it et~al.\/}, {\it Science Advances\/} {\bf 4} (2018).

\bibitem{Mills2018}
A.~K. Mills, {\it et~al.\/}, {\it arXiv\/} {\bf 1902.05997v1} (2018).

\bibitem{Sentef2013}
M.~Sentef, {\it et~al.\/}, {\it Phys. Rev. X\/} {\bf 3}, 041033 (2013).

\bibitem{Sohier2016}
T.~Sohier, {Electrons and phonons in graphene : electron-phonon coupling ,
  screening and transport in the field effect setup}, Ph.D. thesis, De
  L'Universit{\'{e}} Pierre et Marie Curie Sp{\'{e}}cialit{\'{e}} (2016).

\bibitem{Zhou2007}
S.~Y. Zhou, {\it et~al.\/}, {\it Nat. Mater.\/} {\bf 6}, 770 (2007).

\bibitem{Wang2008}
Y.~Y. Wang, {\it et~al.\/}, {\it J. Phys. Chem. C\/} {\bf 112}, 10637 (2008).

\bibitem{Allard2010}
A.~Allard, L.~Wirtz, {\it Nano Letters\/} {\bf 10}, 4335 (2010).

\bibitem{Slonczewski1958}
J.~C. Slonczewski, P.~R. Weiss, {\it Phys. Rev.\/} {\bf 109}, 272 (1958).

\bibitem{Gruneis2008}
A.~Gr{\"{u}}neis, {\it et~al.\/}, {\it Phys. Rev. Lett.\/} {\bf 100}, 037601
  (2008).

\bibitem{Cheng2015}
C.~M. Cheng, {\it et~al.\/}, {\it Appl. Surf. Sci.\/} {\bf 354}, 229 (2015).

\bibitem{Piscanec}
S.~Piscanec, M.~Lazzeri, F.~Mauri, A.~C. Ferrari, J.~Robertson, {\it Phys. Rev.
  Lett.\/} {\bf 93} (2004).

\bibitem{Zhou2006}
S.~Y. Zhou, G.~H. Gweon, A.~Lanzara, {\it Ann. Phys.\/} {\bf 321}, 1730 (2006).

\bibitem{Lazzeri2008}
M.~Lazzeri, C.~Attaccalite, L.~Wirtz, F.~Mauri, {\it Phys. Rev. B\/} {\bf 78},
  081406 (2008).

\bibitem{Chatelain2014}
R.~P. Chatelain, V.~R. Morrison, B.~L.~M. Klarenaar, B.~J. Siwick, {\it Phys.
  Rev. Lett.\/} {\bf 113}, 235502 (2014).

\bibitem{Stern2018}
M.~J. Stern, {\it et~al.\/}, {\it Phys. Rev. B\/} {\bf 97}, 165416 (2018).

\bibitem{Gierz2011}
I.~Gierz, J.~Henk, H.~H{\"{o}}chst, C.~R. Ast, K.~Kern, {\it Phys. Rev. B\/}
  {\bf 83}, 121408(R) (2011).

\bibitem{Liu2011}
Y.~Liu, G.~Bian, T.~Miller, T.~C. Chiang, {\it Phys. Rev. Lett.\/} {\bf 107},
  166803 (2011).

\bibitem{Ulstrup}
S.~Ulstrup, {\it et~al.\/}, {\it J. Phys. Condens. Matter\/} {\bf 27}, 164206
  (2015).

\bibitem{Xu1996}
X.~S., {\it et~al.\/}, {\it Physical Review Letters\/} {\bf 76}, 483  (1996).

\bibitem{Narang2017}
P.~Narang, L.~Zhao, S.~Claybrook, R.~Sundararaman, {\it Adv. Opt. Mater.\/}
  {\bf 5}, 1600914 (2017).

\bibitem{Johannsen2013}
J.~C. Johannsen, {\it et~al.\/}, {\it Phys. Rev. Lett.\/} {\bf 111}, 027403
  (2013).

\bibitem{Butscher2007}
S.~Butscher, F.~Milde, M.~Hirtschulz, E.~Mali{\'{c}}, A.~Knorr, {\it Appl.
  Phys. Lett.\/} {\bf 91}, 203103 (2007).

\bibitem{Sugawara}
K.~Sugawara, T.~Sato, S.~Souma, T.~Takahashi, H.~Suematsu, {\it Phys. Rev.
  Lett.\/} {\bf 98} (2007).

\bibitem{Bostwick2007}
A.~Bostwick, T.~Ohta, T.~Seyller, K.~Horn, E.~Rotenberg, {\it Nat. Phys.\/}
  {\bf 3}, 36 (2007).

\bibitem{Leem2008}
C.~S. Leem, {\it et~al.\/}, {\it Phys. Rev. Lett.\/} {\bf 100} (2008).

\bibitem{Ulstrup2012}
S.~Ulstrup, {\it et~al.\/}, {\it Phys. Rev. B\/} {\bf 86}, 161402(R) (2012).

\bibitem{Joucken2016}
F.~Joucken, {\it et~al.\/}, {\it Phys. Rev. B\/} {\bf 93}, 241101 (2016).

\end{thebibliography}

\end{document}